\pdfoutput=1 
\documentclass{JINST}

\usepackage{amsmath}
\usepackage{amssymb}
\usepackage{subeqnarray}
\usepackage{graphicx}

\usepackage{booktabs} 
\usepackage{array} 
\usepackage{paralist} 
\usepackage{verbatim} 
\usepackage{subfig} 
\usepackage{xcolor}
\let\normalcolor\relax

\usepackage{gensymb} 
\usepackage{amsfonts} 
\usepackage{listings} 
\usepackage[section]{placeins}
\usepackage{microtype} 
\usepackage{appendix} 
\usepackage{url} 
\usepackage{multirow} 

\newcolumntype{C}[1]{>{\centering\arraybackslash}m{#1}}

\title{Low-background temperature sensors fabricated on parylene substrates}

\author{A. Dhar$^a$, J.C. Loach$^{a,b}$, P.J. Barton$^a$, J.T. Larsen$^c$ and A.W.P. Poon$^a$\\
\llap{$^a$}Nuclear Science Division \& Institute for Nuclear and Particle Astrophysics, Lawrence Berkeley National Laboratory, Berkeley, CA 94720, USA.\\
\llap{$^b$}INPAC and Dept. of Physics, Shanghai Jiao Tong University, Shanghai 200240, P. R. China.\\
\llap{$^c$}Earth Sciences Division, Lawrence Berkeley National Laboratory, Berkeley, CA 94720, USA.\\
E-mail: \email{james.loach@sjtu.edu.cn}}

\abstract{Temperature sensors fabricated from ultra-low radioactivity materials have been developed for low-background experiments searching for neutrinoless double-beta decay and the interactions of WIMP dark matter. The sensors consist of electrical traces photolithographically-patterned onto substrates of vapor-deposited parylene. They are demonstrated to function as expected, to do so reliably and robustly, and to be highly radio-pure. This work is a proof-of-concept study of a technology that can be applied to broad class of electronic circuits used in low-background experiments.}


\begin{document}

\maketitle

\setcounter{secnumdepth}{2}

\section{Introduction}

Experiments searching for neutrinoless double-beta decay (0$\nu\beta\beta$) \cite{mjd2,exo200} or the interactions of WIMP dark matter \cite{lux} use a wide array of different technologies. But regardless of their approach these experiments have certain things in common, one of which is a critical need to control radioactive backgrounds. This is achieved, in part, by minimizing the concentrations of radioisotopes in the materials from which the experiment is constructed, from the large homogenous masses that typically comprise the active regions, down to the small components such as sensors and cables that are needed to read out the signals. The complexity of the items in this latter class, the small components, often makes them very difficult to make radio-pure and thus gives them an importance out of proportion to their size. They are a particular issue in experiments whose sensitive volumes are not homogeneous, but rather composed of arrays of discrete units. In these experiments, such as $0\nu\beta\beta$ searches using crystals of germanium, the sensors and cables penetrate deeply into the sensitive region and the effects of the radioactive contaminants they bring with them cannot easily be removed by spatial analysis cuts.

This work is a proof of concept study of a technology that promises to reduce the radioactive contamination in a large category of small components - electronic readout circuits - which must often be located close to the active regions and which are notorious sources of background events. They are notorious because their complexity makes them liable to contamination during production and because many electronic components, such as resistors and substrates, are often made from materials that are highly radio-impure. In this work one of the most basic electronic sensors, the resistive temperature sensor, is manufactured using a substrate of vapor-deposited parylene and traces of sputtered and photolithographically-patterned gold and germanium. The resulting sensors are extremely lightweight, flexible and mechanically robust. They are not the first sensors built using these general technologies \cite{ieee, barton} but the combination of techniques is unique as is the demonstration of the high radiopurity of parylene-based circuits.

\begin{figure}
\centering
\includegraphics{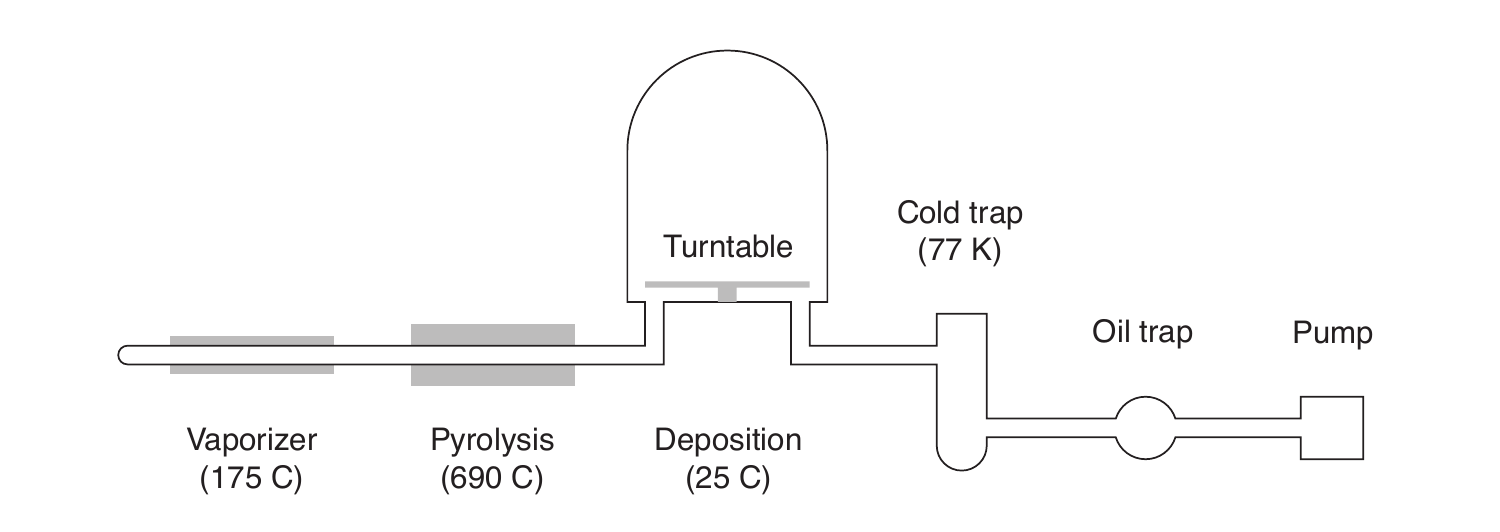}
\caption{The parylene deposition process. Vaporizer and pyrolysis temperatures are for parylene C. The particular parylene machine used in this study was SCS PDS 2010 Labcoter 2, without an oil trap installed.}
\label{fig:parylene}
\end{figure}

\section{Parylene}

The most important physical characteristic of the circuits demonstrated during this work is their use of parylene. Parylene refers to a class of vapor-deposited poly(p-xylylene) polymers with mechanical and electrical properties that are suitable for a wide variety of applications. Parylene is deposited in thin conformal coatings that are pin-hole free, strong and chemical-resistant. These films can be used as insulators, protective layers, dielectrics and substrates. The source material has a high chemical purity and the vapor deposition process by which it is laid down tends to purify it further \cite{parylenepurity}.

The deposition process is illustrated in Figure \ref{fig:parylene}. It begins with the sublimation of the material from its crystalline dimer form, followed by the pyrolization of the vapor into a highly reactive monomer form. The monomer gas then enters the sample chamber, held at room temperature and a pressure of $\sim$0.03\,mbar, where it deposits on all exposed surfaces. The sample typically rotates on a turntable to help ensure uniform coating. The vacuum in the chamber is maintained by an oil pump, which can be separated from the chamber by a cold trap that prevents parylene vapor entering and damaging the pump. An oil trap can be placed between the cold trap and pump to prevent back-flow of oil vapor into the chamber and contamination of the film.

Parylene plays two roles in this work: as a substrate and as a protective, electrically-insulating coating. An initial film of parylene is laid down as a substrate. The parylene is then sputtered with metals that are photolithographically-patterned to create traces. Readout wires are connected and the whole circuit coated with a second layer of parylene.

\section{Sensor fabrication}

\subsection{Design}

\begin{figure}
\centering
\includegraphics{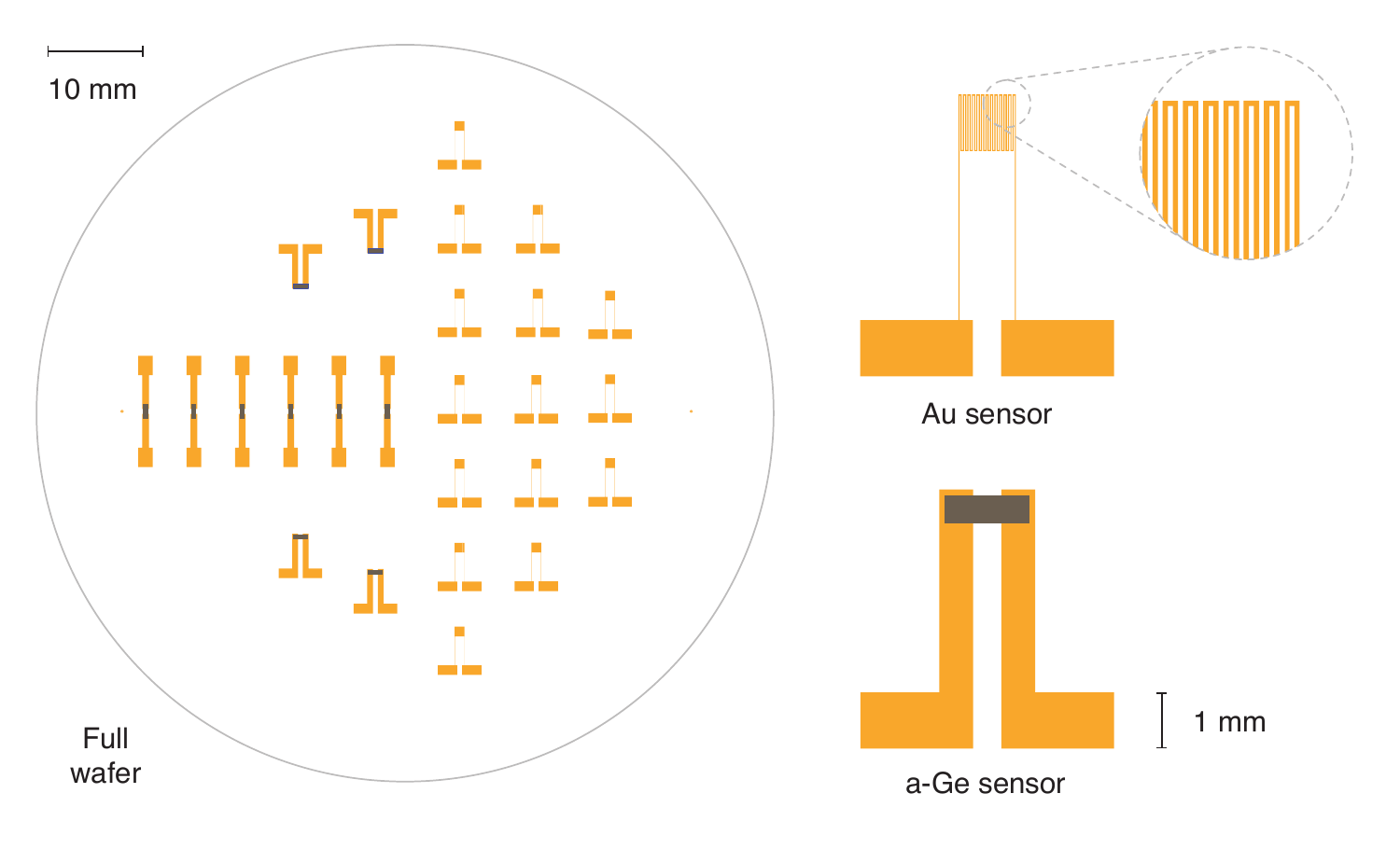}
\caption{Photolithographic masks for a full wafer of sensors (left) and for individual sensors (right). The long vertical elements on the full wafer are features for testing.}
\label{fig:gemask}
\end{figure}

Two kinds of resistive temperature sensor were manufactured in this study: one kind - the \textit{gold} sensors - used a fine serpentine gold trace as the resistive element  and another - the \textit{germanium} sensors - used a resistive element made of amorphous germanium (a-Ge). 

Gold has a resistivity whose dependence on temperature is characteristic of conductive metals in being approximately linear \cite{gold}. Amorphous germanium, by contrast, is a semiconductor and has a resistivity that is much more strongly dependent on temperature. The magnitude of its resistivity is also strongly dependent on conditions under which it is sputtered and on the subsequent chemical processing \cite{barton}. In general, using amorphous germanium results in sensors that are more variable than those made with gold, but that are much more sensitive.

The two sensor designs are illustrated in Figure \ref{fig:gemask}. A typical gold sensor element had serpentine traces 20\,$\mu$m wide, $30\,$mm long and 1\,$\mu$m thick, giving a room temperature resistance of $\sim100\,\Omega$. Germanium sensor elements had traces 0.5\,mm wide, 1.5\,mm long and 1\,$\mu$m thick, giving a typical room temperature resistance of $\sim 12\,{\rm M}\Omega$. On the germanium sensors the traces leading from the connection pads to the resistive element were made thicker than on the gold sensors. On the gold sensors they were kept thin to keep the resistance high but this was not an important consideration for the germanium sensors where the gold contributed negligibly the total resistance. The traces could therefore be made wider to ensure good electrical connections between the gold and germanium, and to make them more robust.

\begin{figure}
\centering
\includegraphics{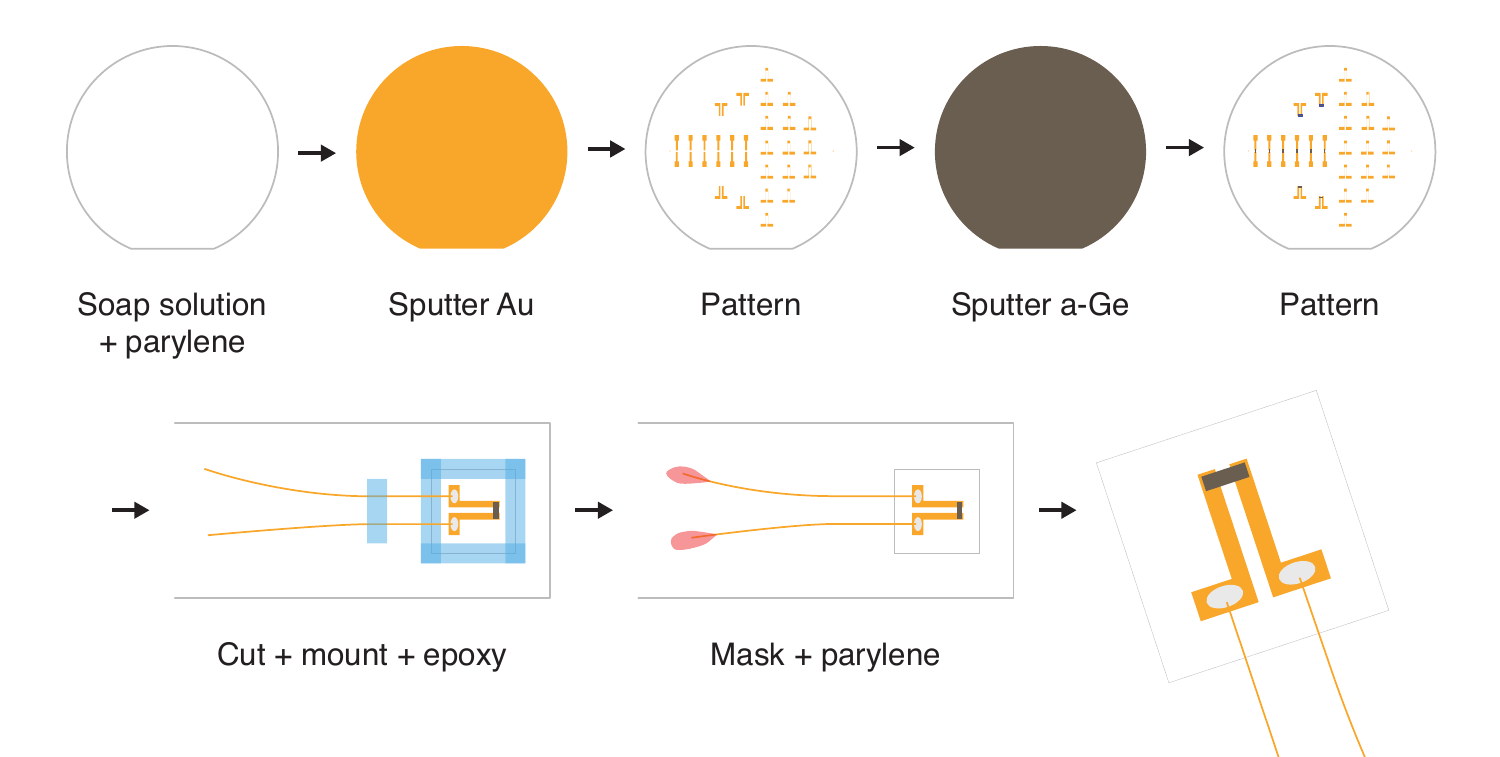}
\caption{Overview of the manufacturing process.}
\label{fig:manufacturing}
\end{figure}

\subsection{Manufacture}

The manufacturing process is illustrated in Figure \ref{fig:manufacturing}. The first step was deposition of the parylene substrate using an SCS PDS 2010 Labcoter 2 machine. The film was deposited onto optically flat (20/10 surface quality) 76 mm diameter fused silica handle wafers, which were used to keep the film flat and secure during subsequent processing. Prior to deposition the wafers were cleaned using methanol, coated with Micro-90 soap solution, and attached to circular aluminum mounting fixtures. The soap solution was used as a release agent to prevent bonding of the parylene to the fused silica. The mounting fixtures were used to allow the parylene to wrap around underneath the edges of the wafers, helping it to remain securely in position during subsequent processing. The wafers were coated with 12\,$\mu$m of parylene C.

\begin{figure}
\centering
\includegraphics{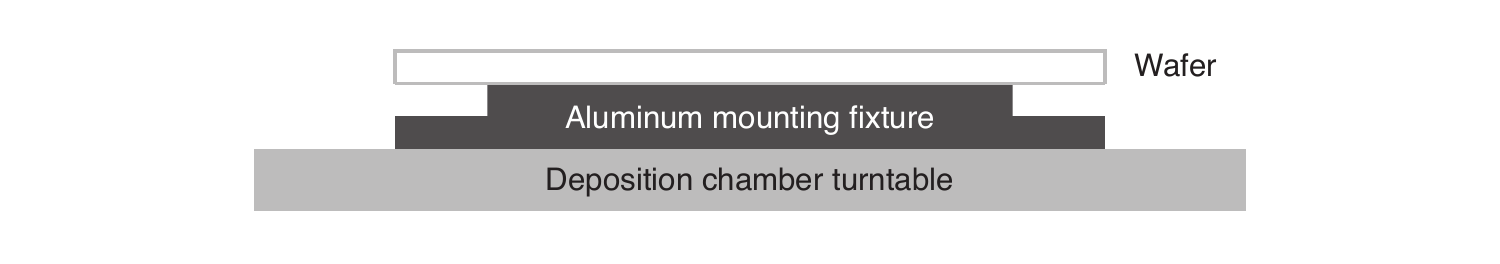}
\caption{Aluminum mounting fixture for fused silica wafers.}
\label{fig:fixture}
\end{figure}

Wafers were detached from their fixtures and sputtered with 15\,nm of titanium followed by 1\,$\mu$m of gold. The titanium was used as bonding agent to help the gold adhere to parylene. Wafers were then patterned and etched using standard photolithographic techniques to create the gold traces: the wafers were dehydrated on a hot plate before being spun with photoresist and exposed to UV light; aqua regia was used to dissolve the gold layer and a diluted mixture of H$_2$O$_2$ and HF used to remove the titanium layer. Wafers were then sputtered with 1\,$\mu$m of germanium, which was then patterned and etched similarly to give the germanium traces.

\begin{figure}
\centering
\includegraphics{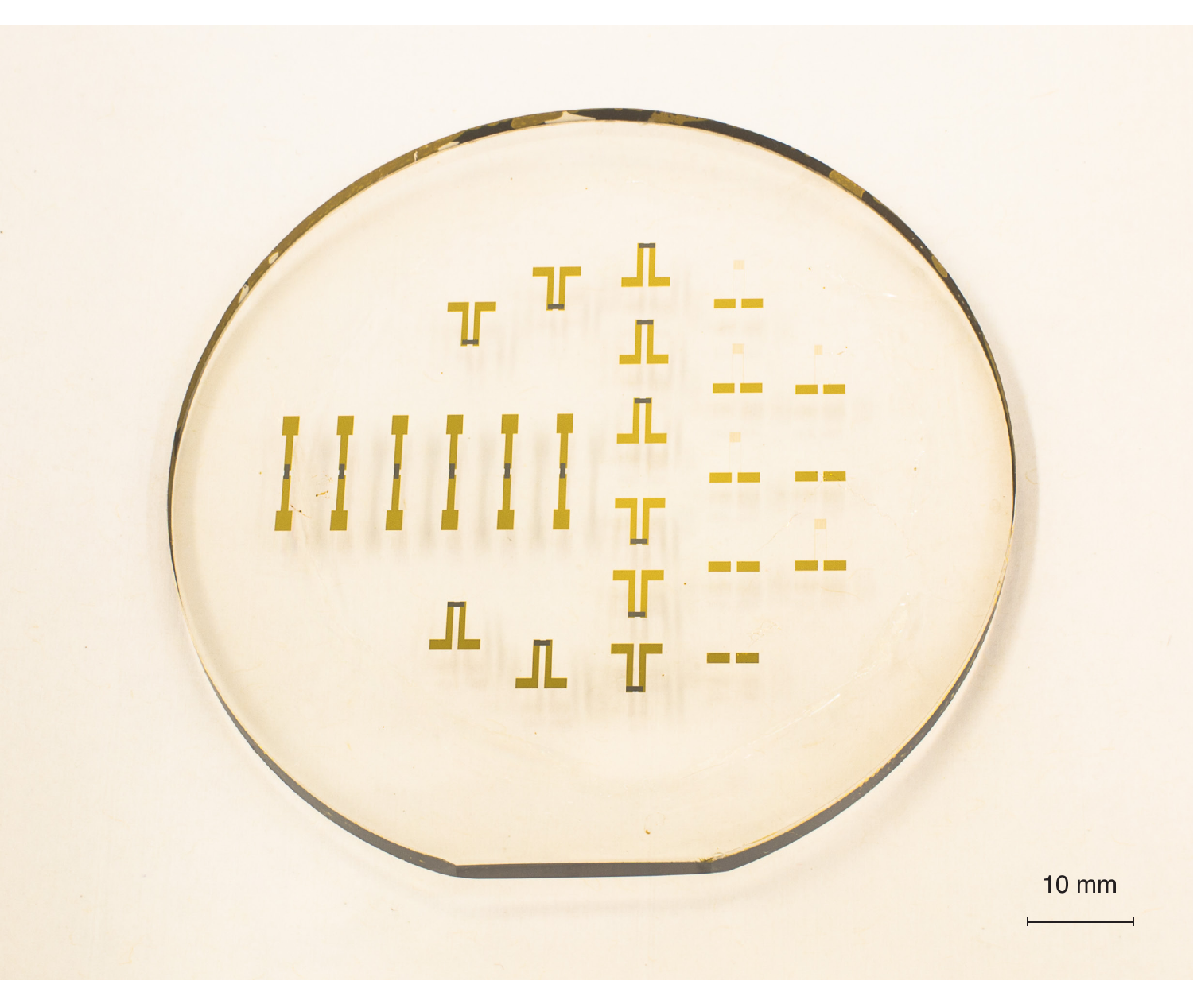}
\caption{A patterned wafer showing germanium sensors and test features on the left, and gold sensors on the right.}
\label{fig:wafer}
\end{figure}

Sensors were then individually removed from the wafers and prepared for attachment of the readout wires. Sensors were cut from the wafer surface, cleaned with methanol and then attached to glass microscope slides with low-tack dicing tape. $76\,\mu$m diameter copper wires, one for each pad, were held in position with dicing tape and then attached to the pads using Hysol TRA-DUCT 2902 silver epoxy. The far ends of the wires were attached to the slide using peelable latex solder masking fluid. Once dry, all dicing tape was removed. The sensor and wires were then bent back away from the surface of the slide and the whole arrangement coated with a second layer of parylene. The masking fluid provided mechanical support, allowing the whole unmasked area to be coated, and also prevented coating of the ends of the wires which needed to be available for making further electrical connections. After the final coating the masking fluid was removed by cutting and peeling.

\subsection{Materials}

The materials used during sensor manufacture are listed in Table \ref{tab:assay}, along with contamination levels and activities for U-238 and Th-232. Contamination levels were measured using inductively-coupled mass spectroscopy (ICP-MS) and direct gamma counting using HPGe detectors. Measurements were not made of the contamination in finished sensors, though this is technically possible using ICP-MS and would be instructive as they would indicate the amount of contamination introduced (or removed) during the production process. However, other measurements made by the authors in the context of the \textsc{Majorana Demonstrator} $0\nu\beta\beta$ experiment indicate that only small amounts of extra contamination, if any, are introduced during parylene deposition and photolithography.

\begin{table}
  \centering

  \caption{Concentrations of $^{232}$Th and $^{238}$U in raw materials. Wires were taken to be 30 mm in length. Total sensor mass was $\sim4$ mg.}
    \begin{tabular}{lrrrrr}
\toprule
    Item  & \multicolumn{1}{c}{Mass \%} &\multicolumn{2}{c}{Conc. (ppb)}  &\multicolumn{2}{c}{Activity (nBq)} \\
\cmidrule(r){3-4} \cmidrule(r){5-6}
              &       & Th-232 & U-238&Th-232 & U-238\\
\midrule
\multicolumn{6}{c}{Au sensors}\\
\midrule
    Copper wire 		   &  $32.0$           & $<0.087$    &  $<0.040$               & $<0.431$       & $<0.608$ \\ 
    Silver epoxy 		   &  $12.2$           & $<0.079$    &   $<0.011$              & $<0.150$      & $<0.064$ \\ 
    Parylene C (sensor)	   &  $51.9$           & $0.53(3)$    &   $ 0.25(6)$              & $4.3(0.2)$      & $6.2(1.5)$     \\ 
    Parylene C (wires)            &  $1.6$             & $0.53(3)$    &   $ 0.25(6)$              & $0.13(0.01)$  & $0.19(0.05)$   \\ 
    Micro-90                         & $\sim0$          &   $<1.5$     &   $<0.6$                   &  $\sim0$       & $\sim0$  \\ 
    Au traces	  	            & 2.3                    &  $47.4(1.1)$ & $2.0(0.4)$              & $17.0(0.4)$   & $2.2(0.4)$ \\ 
    Ti traces	  		  & $\sim0$            & $ <0.4  $  & $<0.1$                     & $\sim0$         & $\sim0$ \\ 
\midrule
     Total                             &  $100.0$             &                &                                 & $<21.9$           & $<9.2$ \\
\midrule
\multicolumn{6}{c}{a-Ge sensors}\\
\midrule
    Copper wire 		   &     $32.3$      & $<0.087$    &  $<0.040$               & $<0.431$         & $<0.608$       \\ 
    Silver epoxy 		   &     $12.4$      & $<0.079$    &   $<0.011$              & $<0.150$         & $<0.064$       \\ 
    Parylene C (sensor)	   &     $52.4$      & $0.53(3)$    &   $ 0.25(6)$              & $4.3(0.2)$        & $6.2(1.5)$       \\ 
    Parylene C (wires)            &     $1.6$        & $0.53(3)$    &   $ 0.25(6)$              & $0.13(0.01)$    & $0.19(0.05)$   \\ 
    Micro-90                         &     $\sim0$    &   $<1.5$     &   $<0.6$                   &  $\sim0$          & $\sim0$          \\ 
    Au traces                         &     $1.1$       & $47.4(1.1)$ & $2.0(0.4)$                 & $8.0(0.2)$        & $1.0(0.2)$       \\ 
    Ti traces                          &     $\sim0$   &  $ <0.4  $  & $<0.1$                       & $\sim0$           & $\sim0$           \\ 
    Ge traces                         &     $0.2$       &   $2.4(0.7)$ & $1.7(0.4)$                 & $0.08(0.02)$    & $0.17(0.04)$   \\  
\midrule
   Total                                &     $100.0$  &                      &                                 & $<13.1$               & $<8.2$ \\

\bottomrule
    \end{tabular}
  \label{tab:assay}
\end{table}

\section{Characterization and performance}

\subsection{Mechanical}

\begin{figure}
\centering
\includegraphics{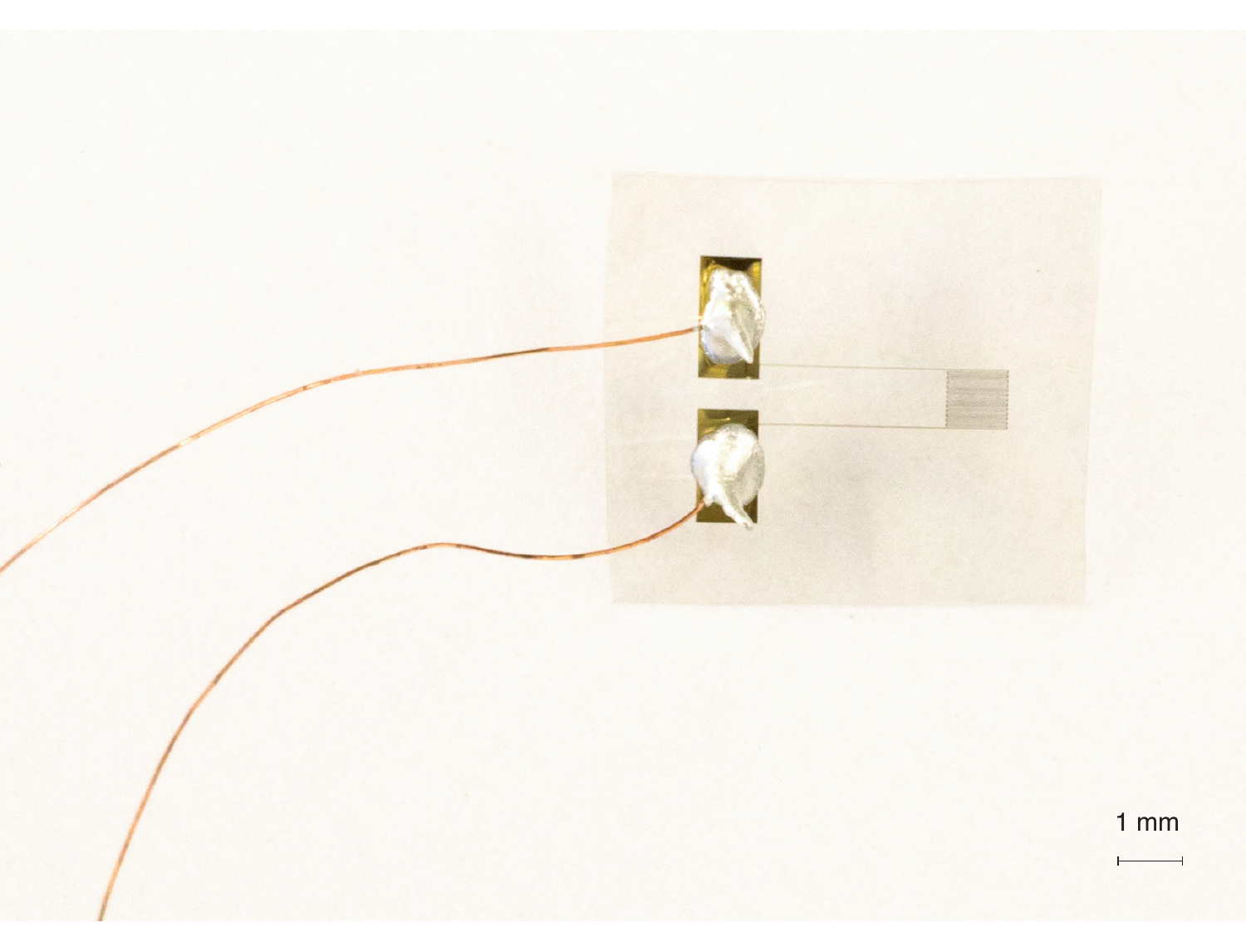}
\caption{A gold-on-parylene resistive temperature sensor.}
\label{fig:finalsensor}
\end{figure}

A completed gold sensor is shown in Figure \ref{fig:finalsensor}. Completed sensors were found to be mechanically robust and able to sustain substantial amounts of handling during testing. They withstood sustained temperatures of 200$^{\circ}$\,C as well as rapid immersion in water and liquid nitrogen. The only mechanical weak point was found to be where the wires entered the silver epoxy, at which point the wires were prone to mechanical fatigue from repeated bending.

Some failure modes were observed during sensor manufacture, mostly related to the fine traces on the gold sensors being scratched prior to the application of the second, protective layer of parylene. This issue can be easily remedied by using larger trace widths.

\subsection{Electrical}

Gold sensors were tested on a custom-built apparatus using a two-lead sensing method. Sensors were clamped to a copper rod, which was rapidly cooled to $<100\,$K and then allowed to warm up slowly in air. Data were recorded during the warm up using a Lakeshore 325 temperature controller. Testing of germanium sensors proceeded similarly but in the more controlled environment of a variable-temperature vacuum cryostat.

\begin{figure}
\centering
\includegraphics{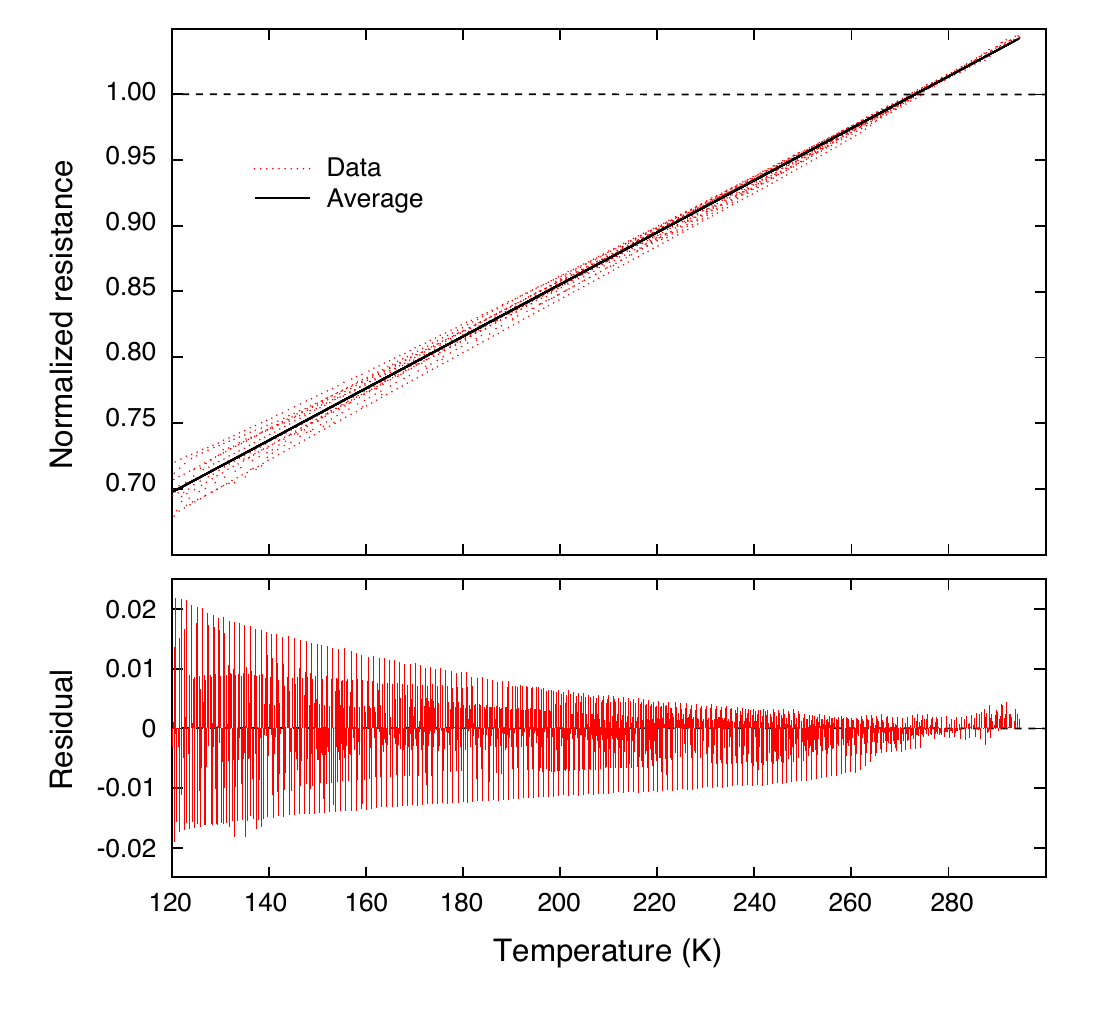}
\caption{Variation of resistance with temperature for a random selection of gold sensors normalized to the average resistance at 273 K (top) and the variation compared to the average (bottom).}
\label{fig:goldresults}
\end{figure}

\begin{figure}
\centering
\includegraphics{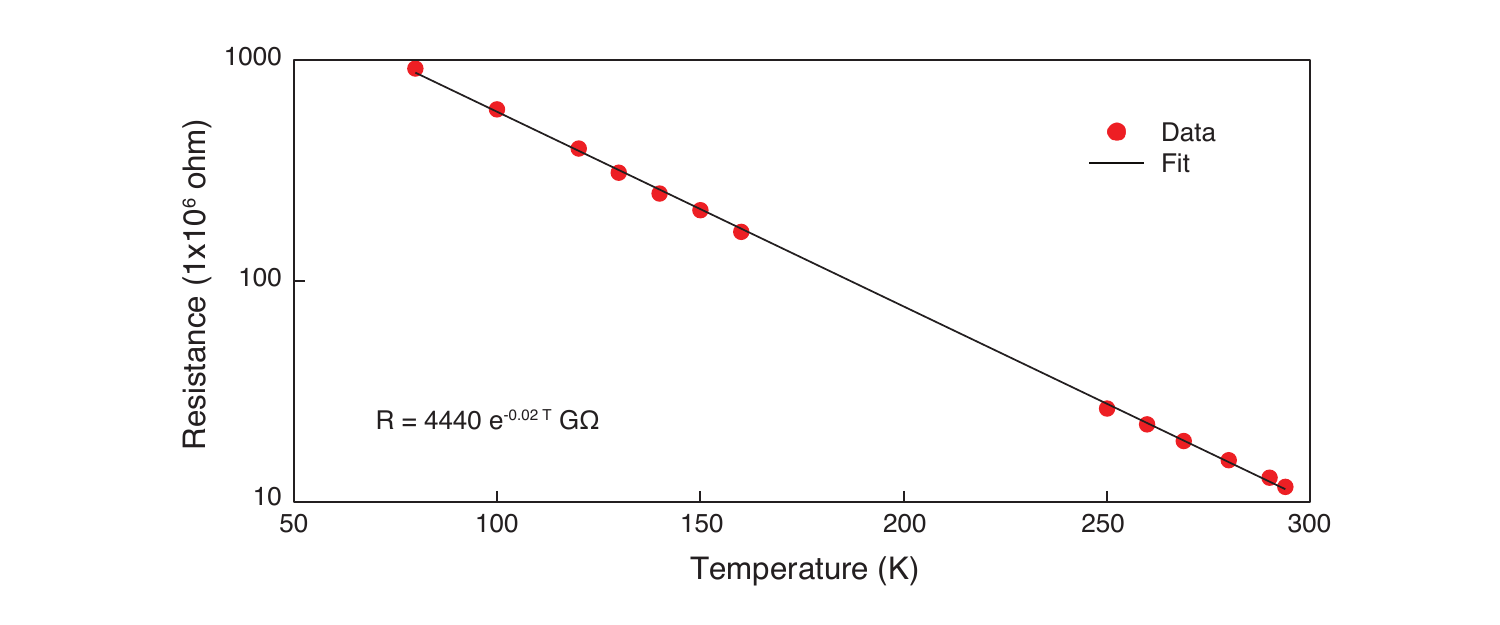}
\caption{Variation of resistance with temperature for a typical germanium sensor.}
\label{fig:germaniumresults}
\end{figure}

Figure \ref{fig:goldresults} shows results from a random selection of gold sensors and Figure \ref{fig:germaniumresults} shows data for a single germanium sensor. Gold sensors showed a strongly linear dependence of the resistance on temperature with small variations in slope from sensor to sensor. The observed variation in slope can be attributed to sensor-to-sensor and wafer-to-wafer variations in the etching process. Fewer germanium sensors were tested but the observed temperature dependence of the resistance was consistent with expectations. Germanium test features revealed a dependence of resistance on position on the wafer that was pronounced at radii near the periphery. This behavior can be attributed to non-uniformities in germanium sputtering and etching.

No measurements were performed to establish the stability of resistances over time. Such measurements would be interesting for germanium sensors because the resistances of germanium films are known to be sensitive to the conditions under which they are deposited \cite{barton} and they are expected to be similarly sensitive to storage conditions and subsequent handling. The latter two effects might be mitigated by parylene coating, given the excellent barrier properties and chemical inertness exhibited by the material.

\section{Conclusions}

\begin{figure}
\centering
\includegraphics{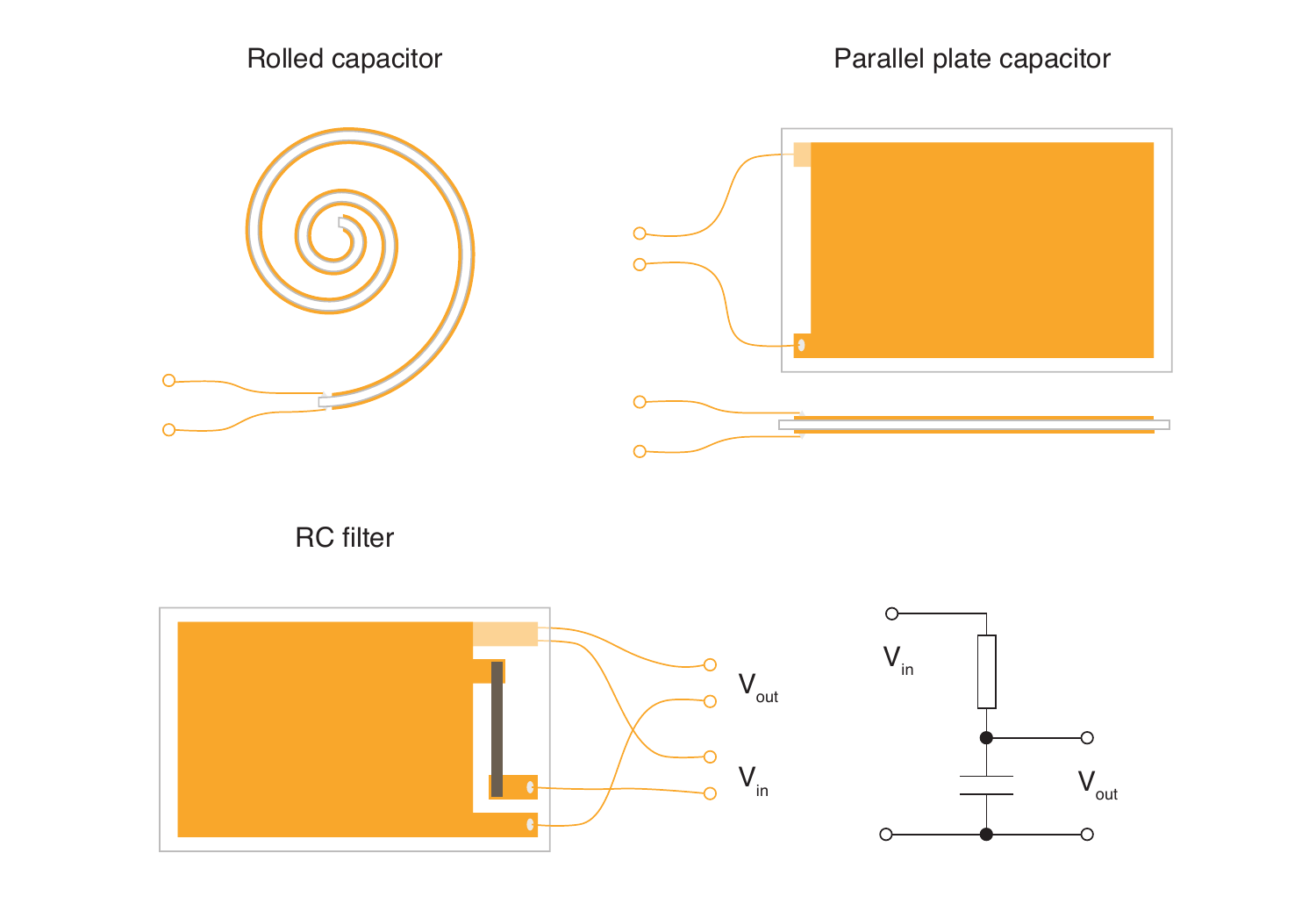}
\caption{Designs for capacitors and an RC filter. For clarity these illustrations omit the outer, protective layer of parylene that could be applied to each circuit.}
\label{fig:cap}
\end{figure}

The temperature sensors described in this work performed as expected, both mechanically and electrically, and have been found to be highly radio-pure. In their current form they are suitable for use in low-background experiments. The different resistance profiles of the two types of temperature sensor make them suitable for use in slightly different temperature regimes; in particular, gold sensors can likely be used at lower temperatures than are appropriate for the higher-resistance germanium sensors. Minor design changes for a second batch of sensors would be removing surplus gold from the connection pads, making the distance between pads and sensor elements longer (to give more flexibility in positioning) and increasing the widths of the pad-to-sensor traces on the gold sensors (to reduce the potential for damage during processing).

Parylene substrates are also suitable for producing a variety of other circuits and some examples are illustrated in Figure \ref{fig:cap}. The material is well-suited to the role of a capacitor dielectric and substantial capacitances can be generated using large-area films of sub-micron thickness. These thin-film capacitors can be rolled before application of the second parylene layer to make them mechanically stable in their final state. In addition, amorphous germanium resistors can be incorporated to make RC filters. Parylene can also be used to coat capacitors made using other ultra-clean dielectrics such as fused silica, on which gold and amorphous germanium can be similarly deposited and patterned. Here the parylene would provide mechanical protection for the traces and stabilization for the amorphous germanium resistors.

\section{Acknowledgements}

This work was supported by the U.S. Department of Energy, Office of Science, Office of Nuclear Physics, under Contract No. DE-AC02-05CH11231 and by the Shanghai Key Lab for Particle Physics and Cosmology (SKLPPC), Grant No. 15DZ2272100.

\bibliography{tempsensor}{}
\bibliographystyle{ieeetr}

\end{document}